\documentclass[12pt]{iopart}

\usepackage{latexsym,epsfig, amsfonts}

\expandafter\let\csname equation*\endcsname\relax
\expandafter\let\csname endequation*\endcsname\relax
\usepackage{amsmath}
\usepackage{bm}

\def\egr{{\bf e} }

\def\kgr{{\bf k} }

\def\0gr{{\bf 0} }

\begin{document}

\title{Phase diagram and correlation functions of the anisotropic imperfect Bose gas in $d$ dimensions}

\author{P. Jakubczyk}
\address{Institute of Theoretical Physics, Faculty of Physics, University of Warsaw, Pasteura 5, 02-093 Warsaw, Poland
}
\address{Max Planck Institute for Solid State Research, Heisenbergstr. 1, 70569, Stuttgart, Germany}
\ead{pawel.jakubczyk@fuw.edu.pl}

\author{J. Wojtkiewicz}
\address{Department of Mathematical Physics, Faculty of Physics, University of Warsaw, Pasteura 5, 02-093 Warsaw, Poland}
\ead{jacek.wojtkiewicz@fuw.edu.pl}



\begin{abstract}
We study an anisotropic variant of the $d$-dimensional imperfect Bose gas, where the asymptotic behaviour of the dispersion $\epsilon_{\bf k}$ at vanishing momentum $\bf{k}$ may differ from the standard quadratic form. The analysis reveals the key role of the shift exponent $\psi$ governing the asymptotic behaviour of the critical temperature $T_c(\mu)$ as a function of the chemical potential $\mu$ at $T_c\to 0$. We argue that the universality classes of Bose-Einstein condensation admitted by the model may be classified according to the allowed values of $\psi$ so that spatial dimensionality has only an indirect impact on the transition properties. We analyse the correlation function of the model and discuss its asymptotics depending on the direction. Both for the perfect and imperfect anisotropic Bose gases, the correlation function $\chi({\bf x})$ at $T>T_c$ turns out to show either exponential decay or exponentially damped oscillatory behaviour depending on the orientation of ${\bf x}$ with respect to the dispersion anisotropies.

\end{abstract}
\pacs{05.30.Jp, 03.75.Hh, 64.60.F-}

\maketitle

\section{Introduction} 
Bose gases are among the few systems that have remained within central focus of condensed matter physics for many decades. Certainly the detailed context of these studies evolved over time. In particular, the last years witnessed remarkable progress in experimental control over ultra-cold atomic gases (bosons in particular) in optical lattices \cite{Greiner02, Bloch08}. An accurate theoretical \cite{Jaksch98} description of such systems necessarily involves a lattice and inter-particle interactions. For the majority of theoretical models, such as the Bose-Hubbard model \cite{Fisher89}, these may be treated only approximately, and, in most cases, involve costly numerical techniques. Recognised theoretical approaches to interacting Bose systems on a lattice include, among others, strong-coupling expansions \cite{Sengupta05, FFU06},
reflection positivity \cite{ALSSY}, renormalisation group \cite{Dupuis09, Sinner09, Sinner10, Rancon12, Rancon13} and bosonic dynamical mean field theory \cite{Byczuk08, Panas15}.

Bearing the above in mind it seems worthwhile investing some effort in development of simplified theoretical models which are susceptible to an exact analytical treatment.  
In the present paper we adapt a model known as the imperfect Bose gas (IBG) \cite{Davies72, Buffet83, Berg84, Zagrebnov01, Napiorkowski11} to the anisotropic ($d$-dimensional) case. This may result from an underlying  lattice. We show how tuning the lattice parameters may lead to varying the asymptotic behaviour of the dispersion $\epsilon_{\bf k}$, which in turn alters the universality class of the transition as well as the lower and upper critical dimensions as compared to the standard isotropic gas in the continuum. The present model is exactly solvable and allows for tracking the emergence of universality without a recourse to renormalisation group theory. The anisotropies of the dispersion are also a source of interesting behaviour of the correlation function $\chi({\bf x})$ which, as we demonstrate, may (in the normal phase) show either monotonous exponential decay, or exponentially damped oscillations, depending on the direction of ${\bf x}$. This  property occurs both for the perfect (noninteracting) and imperfect Bose gases.    

The outline of the paper is as follows: 
In Sec. 2 we introduce the model together with the saddle-point technique leading to its exact solution. Of high relevance is the asymptotic form of the dispersion  $\epsilon_{\bf k}$, which we relate to specific lattice models. Sec. 3 contains a presentation of our results concerning the phase diagram and properties of the Bose-Einstein condensation. These are compared to the case of the noninteracting Bose gas \cite{Huang, Ziff77} in the subsequent Sec. 4. We analyse the correlation function of both the perfect and imperfect Bose gases in Sec. 5 with emphasis on the asymptotic features and properties related to the anisotropy.  We summarise the paper in Sec. 6.   

\section{Model and the saddle-point method} 
 We consider a system of spinless bosons at a fixed temperature $T$, chemical potential $\mu$ and volume $V=L^d$. The system is governed by the Hamiltonian 
\begin{equation}
 \hat{H}=\sum_{\bf k} \epsilon_{\bf k}\hat{n}_{\bf k}+\frac{a}{2V}\hat{N}^2 \;.
\label{Hamiltonian}
\end{equation} 
We impose periodic boundary conditions and the dispersion relation $\epsilon_{\bf k}$ is, for now, unspecified. The ${\bf k}$ summation runs over the first Brillouin zone. We will in particular discuss the hypercubic lattice in Sec.~2.1. The quantity $\hat{N}=\sum_{\bf k}\hat{n}_{\bf k}$ is the total particle number operator. The repulsive mean-field interaction term $H_{mf}=\frac{a}{2V}\hat{N}^2$ ($a>0$) may be derived 
from the long-range repulsive part  $v(r)$ of a 2-particle interaction potential upon performing the Kac scaling limit $\lim_{\gamma\to 0}\gamma^dv(\gamma r)$, i.e. for vanishing interaction strength and diverging range. Therefore, the interaction employed in Eq.~(\ref{Hamiltonian}) corresponds to a limiting situation where the force between two particles is taken sufficiently long ranged and weak so that the actual distance between them becomes less and less relevant. The presence of the $1/V$ factor assures extensivity of the system. The continuum version of the model with $\epsilon_{\bf k}\sim {\bf k}^2$ was studied in Refs.~\cite{Davies72, Buffet83, Berg84, Zagrebnov01, Napiorkowski11, Napiorkowski13, Jakubczyk13, Diehl17}. The presence of the mean-field interaction $H_{mf}$ has a profound impact on the properties of Bose-Einstein condensation and makes it substantially different as compared to the ideal Bose gas \cite{Ziff77}.  

Following the line of Ref.~\cite{Napiorkowski11} the grand canonical partition function $\Xi(T,V,\mu)$ can be represented via the contour integral 
\begin{equation}
\label{GC}
 \Xi (T,V,\mu) = -i e^{\frac{\beta\mu^2}{2a}V}\left(\frac{V}{2\pi a \beta}\right)^{1/2} \int_{\theta\beta-i\infty}^{\theta\beta+i\infty}dse^{-V\phi(s)}\;, 
\end{equation}
where the parameter $\theta<0$ is arbitrary and 
\begin{equation}
\phi(s) = \frac{-s^2}{2a\beta} +\frac{\mu s}{a} - \frac{1}{V}\sum_{{\bf k}\neq 0}\ln \frac{1}{1-e^{s-\beta \epsilon_{\bf k}}}-\frac{1}{V}\ln \frac{1}{1-e^{s}} 
\label{phi}
\end{equation}  
with $\beta = (k_B T)^{-1}$. We isolated the last term and assumed that there is a single minimum of the dispersion $\epsilon_{\bf k}$ in the Brillouin zone, which occurs at ${\bf k} =0$ with $\epsilon_{\bf k=0}=0$. The integration contour in Eq.~(\ref{GC}) may be deformed to pass through a saddle point $\bar{s}$, located in the negative part of the real axis in the corresponding complex plane (see below). The factor $V$ in the exponential in Eq.~(\ref{GC}) then assures that the saddle-point approximation to the integral in Eq.~(\ref{GC}) becomes exact for $V\to\infty$, i.e. 
\begin{equation}
\lim_{V\to\infty}\frac{1}{V}\log \Xi (T,V,\mu) = \frac{\beta\mu^2}{2a}-\phi (\bar{s})\;. 
\label{free_en}
\end{equation}
This reduces the problem of evaluating the integral in Eq.~(\ref{GC}) to solving the saddle-point equation 
\begin{equation} 
\label{saddle-point_eq}
\phi'(\bar{s})=0\; 
\end{equation}
and subsequently computing $\phi (\bar{s})$.
Explicitly Eq.~(\ref{saddle-point_eq}) reads 
\begin{equation}
\label{sp2}
-\frac{\bar{s}}{a\beta}+\frac{\mu}{a}=\frac{1}{V}\sum_{{\bf k}\neq 0} \frac{1}{e^{\beta \epsilon_{\bf k}-\bar{s}}-1}+\frac{1}{V}\frac{1}{e^{-\bar{s}}-1} \;.
\end{equation}
Thus far we adapted the reasoning of Ref.~\cite{Napiorkowski11} to the present, more general case where the dispersion $\epsilon_{\bf k}$ remains unspecified. For the continuum case, where  $\epsilon_{\bf k}\sim {\bf k}^2$ the ${\bf k}$-summation in Eq.~(\ref{sp2}) may be performed for $V\to\infty$ yielding a Bose function. This is not the case here, where even convergence properties determining the lower critical dimension for condensation do depend on $\epsilon_{\bf k}$.   

We now consider the thermodynamic limit $V\to\infty$ where $\sum_{{\bf k}\neq 0}\to\frac{V}{(2\pi)^d}\int d^d k$. The term involving the ${\bf k}$ summation occurring in Eq.~(\ref{sp2}) is then replaced by the integral 
\begin{equation}
\label{I_d}
I_d(\beta, \bar{s})=\frac{1}{(2\pi)^d}\int_{BZ}d^d k\frac{1}{e^{\beta\epsilon_{\bf k}-\bar{s}}-1}\;.
\end{equation}
  In the thermodynamic limit and for $T\leq T_c$ or $T\to T_c^+$ we have $\bar{s}\to 0^-$ (see Sec. 3) and therefore the integral 
is dominated by the vicinity of ${\bf k}=0$. 

In what follows we will specify to the cases where the asymptotic behaviour of the dispersion $\epsilon_{\bf k}$ around ${\bf k}=0$ may be cast in the form 
\begin{equation}
\epsilon_{\bf k} \approx \sum_{i=1}^d t_i |k_i|^{\alpha_i}\;, 
\label{asymptotics}
\end{equation}
where $\alpha_i$ and $t_i$ are positive real numbers for $i\in\{1\dots d\}$. We discuss this restriction and the specific relevant cases below in Sec.~2.1.
\subsection{Genesis of the dispersion asymptotics}
Consider now a system of noninteracting bosons on a hypercubic lattice, subject to periodic boundary conditions. The corresponding Hamiltonian reads 
\begin{equation}
\label{Ham2}
H_{free}=-\sum_{({\bf x}{\bf y})}t_{{\bf x}{\bf y}}c_{{\bf x}}^{\dagger}c_{{\bf y}} = \sum_{{\bf k}} \epsilon_{{\bf k}}c_{{\bf k}}^{\dagger}c_{{\bf k}}\;,
\end{equation}
where ${\bf x}$ and ${\bf y}$ label the lattice points, $({\bf x}{\bf y})$ denotes a pair of points, the hoppings $t_{{\bf x}{\bf y}}$ are assumed to be real, and the second equality follows from diagonalising the Hamiltonian with the Fourier transform. The dispersion is expressed as 
\begin{equation}
\epsilon_{\bf k} = \sum_{\bf x} 2 t_{\bf x}\left(1-\cos\left({\bf k}\cdot{\bf x}\right)\right)+const 
\label{disp}
\end{equation}
with $t_{\bf x} = t_{\bf 0x}$. One can add a constant to the original Hamiltonian to assure that the constant appearing in Eq.~(\ref{disp}) vanishes. For positive hopping coefficients the dispersion is always quadratic around ${\bf k}=0$. This does not have to be the case in situations where the hoppings carry different signs. For example, consider the case 
 \begin{equation}
 \epsilon_\kgr=2t\sum_{i=1}^d (1-\cos k_i)+ 2t_2 (1-\cos 2 k_1).
\label{dispersion_FAF2}
\end{equation}
If the $t, t_2$ couplings fulfil: $t_2=-t/4$, the coefficient of $k_1^2$ vanishes, and
the asymptotics takes the form
 \begin{equation}
 \epsilon_\kgr\sim t \sum_{i=2}^d  k_i^2 + \tau' k_1^4,
\label{asympt_FAF2}
\end{equation}
which is anomalously flat in the direction 1. 

Now generalise the above case by taking $t_{\egr_i} =t>0$ for $i=1,\dots, d$;  $t_{2\egr_1} =t_2$, $t_{3\egr_1} =t_3$, \dots, $t_{m\egr_1} =t_m$. The analysis then shows that by a suitable choice of the hopping constants $t_2, \dots, t_m$ one can achieve vanishing of the derivatives of $\epsilon_k$
in direction 1 up to the order $2m$. The asymptotic form of the dispersion then reads
 \begin{equation}
 \epsilon_\kgr\sim t \sum_{i=2}^d  k_i^2 + \tau_m k_1^{2m}, \quad \tau_m>0\;.
\label{asympt_FAFm}
\end{equation}
Such a construction can be carried out independently in each of the spatial directions, leading to the asymptotics of the form of Eq.~(\ref{asymptotics}). The resulting exponents $2m_i$ are even natural numbers, which may by made arbitrarily large by judicious choices of the hoppings. For the sake of generality we allow the exponents appearing in Eq.~(\ref{asymptotics}) to be arbitrary positive real numbers.  

The asymptotics given by Eq.~(\ref{asymptotics}) displays a minimum at $\kgr=0$ being an {\em isolated point}.
By modifications of certain assumptions of the model, one can obtain a {\em   non isolated} minimum. For instance, if we relax the demand
of strict translational invariance to translational invariance with some period, we can achieve a completely flat dispersion
in one or more directions. Examples of lattices with such dispersions appear in the Hubbard models with 'flat-band ferromagnetism'
\cite{Tasaki}
(the 'sawtooth' one-dimensional lattice, or two-dimensional kagom\'e lattice). The analysis of systems with
non-isolated critical points
('completely flat dispersion') will not be considered here.
\subsection{The `cone' dispersion}
It is clear that the asymptotic formula assumed in Eq.~(\ref{asymptotics}) also applies to the ultrarelativistic Bose gas in continuum, where the relativistic dispersion is dominated by the kinetic contribution, i.e. 
\begin{equation}
\sqrt{m^2c^4+c^2{\bf p}^2}\approx c|{\bf p}| \;.
\end{equation}
Of course, we consider only the kinetic energy term and neglect  QFT effects such as creation of particle-antiparticle pairs etc. 

The model with relativistic dispersion can be viewed as of academic interest only, but let us remember that in $d=2$ the dispersion relation on the hexagonal lattice displays exactly the asymptotic form $\sim |{\bf p}|$ as above. In higher dimensions it is also possible to achieve the 'cone' asymptotic form by a suitable tuning of the hopping constants. 

\section{Phase diagram} 
We now analyse the properties of the model defined by Eq.~(\ref{Hamiltonian}) and assuming the asymptotic behaviour of $\epsilon_{\bf k}$ given by Eq.~(\ref{asymptotics}).

\subsection{Lower critical dimension}
The last term in Eq.~(\ref{sp2}) has the interpretation of the condensate density $n_0$ \cite{Napiorkowski11}. The system finds itself in the low-temperature phase (i.e. the phase hosting the Bose-Einstein condensate) if $n_0$ remains finite in the thermodynamic limit. If the integral $I_d(\beta, \bar{s})$ [see Eq.~(\ref{I_d})] 
 diverges for $\bar{s}=0$, then Eq.~(\ref{sp2}) admits a solution for $\bar{s}<0$ at any $V$, and $\lim_{V\to\infty} \bar{s} < 0$ at any $T>0$. Indeed, the left-hand side (LHS) of Eq.~(\ref{sp2}), viewed as a function of $\bar{s}$, may be represented as a straight line of negative slope. On the other hand, the first term on the right-hand side (RHS) represents (for $\bar{s}<0$) a continuous convex function unbounded from below. If this function diverges for $\bar{s}\to 0^-$, these two curves must intersect at some $\bar{s}<0$. This implies that the last term of  Eq.~(\ref{sp2}) vanishes in the thermodynamic limit, therefore condensation does not occur at any $T>0$, and, in consequence, the system finds itself below the lower critical dimension. Conversely, if the integral in Eq.~(\ref{sp2}) is finite at $\bar{s}=0$, than by lowering $T$ one may shift the straight line corresponding to the LHS of Eq.~(\ref{sp2}) such that the two curves do not intersect. The difference is then compensated by the last term of  Eq.~(\ref{sp2}). For a graphical illustration we refer to Fig.~1 of Ref.~\cite{Napiorkowski11}.

 We therefore seek to derive a condition for the convergence/divergence of the $d$-dimensional integral of Eq.~(\ref{I_d}). The integral is dominated by the vicinity of ${\bf k}=0$ and the replacement of $\epsilon_{\bf k}$ by its asymptotics at small $k$ has no impact on convergence. We use the asymptotics given in Eq.~(\ref{asymptotics}) and perform the variable transformation
\begin{equation}
x_i^2=\beta t_i|k_i|^{\alpha_i} \;\;\;\;\; \textrm{for} \;\;\;i=1\dots d\;. 
\label{transf}
\end{equation}    
This brings the integral to the form 
\begin{equation}
\label{I}
I= \pi^{-d} \int dx_1\dots dx_d \prod_{i=1}^{d} \left[ \frac{x_i^{-1+\frac{2}{\alpha_i}}}{\alpha_i}  \left(\frac{1}{\beta t_i}\right)^{\frac{1}{\alpha_i}}     \right]  \frac{1}{e^{{r_d}^2}-1}\;,
\end{equation}
where 
\begin{equation}
{r_d}^2 = x_1^2+\dots x_d^2\;.
\end{equation} 
The expression for $I$ differs from the integral in Eq.~(\ref{I_d}) only by a finite expression.

We subsequently introduce the spherical coordinates 
\begin{eqnarray}
\label{spherical}
x_i = \left\{ 
\begin{array}{l l}
  r_d \cos\theta_1 & \quad \mbox{$i=1$} \\
  r_d\left(\prod_{j=1}^{i-1}\sin\theta_j \right)  \cos\theta_i & \quad \mbox{$i \in \{2,\dots ,d-1\}$} \\ 
  r_d\left(\prod_{j=1}^{i}\sin\theta_j \right)	 & \quad \mbox{$i=d$}  \quad. \\ \end{array}  \right. 
\end{eqnarray}
The angular integrations then give a constant and the convergence of the integral becomes equivalent to finiteness of the expression 
\begin{equation}
\label{ref_int}
I=\textrm{const} \int_0^\Lambda dr \frac{r^{-1+\frac{2}{\psi}}}{e^{r^2}-1}\;,
\end{equation}
where we replaced $r_d$ with $r$ and introduced 
\begin{equation}
\label{psi}
\frac{1}{\psi} = \sum_{i=1}^{d} \frac{1}{\alpha_i}\;.
\end{equation}
The cutoff $\Lambda>0$ may be taken arbitrarily small without influencing convergence.  The integral of Eq.~(\ref{ref_int}) converges for $\psi<1$ and diverges otherwise. The condition for the lower critical dimension $d_L$ therefore reads 
\begin{equation}
\frac{1}{\psi}=\frac{1}{\alpha_1}+\frac{1}{\alpha_2}+\dots \frac{1}{\alpha_{d_L}} =1\;.
\end{equation} 
We conclude this subsection by stating that condensation occurs if and only if 
\begin{equation} 
\label{psi_nier}
\psi <1\;.
\end{equation} 
In the most standard case where $\alpha_i=2$ for each $i$, we recover $d_L=2$. Obviously, higher values of the exponents $\{\alpha_i\}$ tend to rise $d_L$. Using the observations from Sec. 2.1 we note that $d_L$ may be pushed to arbitrarily high values.  
In the next subsection we demonstrate that $\psi$ may be given the interpretation in terms of the shift exponent, which governs the low temperature asymptotics of the phase boundary $T_c(\mu)$.

\subsection{Critical line} 
If $d>d_L$ and the thermodynamic state is adjusted to be exactly at the phase transition, we have $\bar{s}\to 0^-$, and, in addition, $n_0\to 0$ for $V\to\infty$. Eq.~(\ref{sp2}) then yields the condition for the condensation temperature $T_c$ as a function of the chemical potential (or vice versa).
\begin{equation}
\frac{\mu}{a}=\frac{1}{(2\pi)^d}\int_{BZ} d^d k\frac{1}{e^{\epsilon_{\bf k}/(k_B T_c)}-1}\;. 
\label{T_c_gen}
\end{equation} 
We note that the expression on the RHS of the above equation (with $T_c$ replaced by $T$) coincides with the expression for the critical density $n_c^p$ of a perfect Bose gas at a given temperature $T$ \cite{Ziff77}. We therefore recover the relation 
\begin{equation}
\mu_c(T) = an_c^p(T).
\end{equation}

The dependence of $T_c$ on $\mu$ as given implicitly by Eq.~(\ref{T_c_gen}) is nonuniversal, i.e. depends on the specific form of $\epsilon_{\bf k}$. We note however, that in the limit of $T_c\to 0$ the integral becomes dominated by ${\bf k}\approx 0$. This allows us to replace $\epsilon_{\bf k}$ with its asymptotic form given by Eq.~(\ref{asymptotics}) and investigate the universal asymptotic behaviour of $T_c(\mu)$ at low temperatures. By performing the transformation of Eq.~(\ref{transf}), passing to the spherical coordinates Eq.~(\ref{spherical}), we extract the dependence $\mu (T_c)$ in the form 
\begin{equation}
\mu \simeq \textrm{const} \times T_c^{\frac{1}{\psi}}
\end{equation} 
with $\psi$ given by Eq.~(\ref{psi}). The non-universal prefactor involves the product $\prod_{i=1}^{d}\left[\frac{2}{\alpha_i}\left( \frac{k_B}{t_i} \right)^{\frac{1}{\alpha_i}}  \right]$ and the integral from Eq.~(\ref{I}).
The exponent $\psi$ governing the asymptotic dependence $T_c(\mu)$ is however universal and given by Eq.~(\ref{psi}). For $\epsilon_{\bf k}\sim {\bf k}^2$ it agrees with the known renormalisation-group result for $\phi^4$-type effective field theory, which reads \cite{Sachdev_book, Millis93, Jakubczyk08} 
\begin{equation}
\psi = \frac{z}{d+z-2}\;. 
\label{psi_stand}
\end{equation} 
For the presently relevant case of interacting Bose particles the dynamical exponent $z=2$. In consequence the above formula agrees with Eq.~(\ref{psi}) as long as $\alpha_i = 2$ for each $i\in \{1\dots d\}$. Interestingly, neither the above standard formula, nor our Eq.~(\ref{psi_nier}) allows for the possibility of realising a phase diagram with $\psi>1$.

\subsection{Critical behaviour} 
We now analyse the saddle-point equation Eq.~(\ref{sp2}) asymptotically close to the phase transition, on the high-temperature side,  where $\bar{s}$ can be used as a small parameter. We split the integration domain (Brillouin zone)  into a union of an arbitrary open  neighbourhood $K_d^\Lambda$ of ${\bf k}=0$, which is of a characteristic (arbitrarily small) size $\Lambda$, and the remainder $\mathcal{R}^\Lambda$, so that in Eq.~(\ref{I_d})
\begin{equation}
\int_{BZ} = \int_{K_d^\Lambda} + \int_{\mathcal{R}^\Lambda}\;.
\end{equation}
The contribution to the integral $I_d(\beta, \bar{s})$ in Eq.~(\ref{I_d}) from the remainder region is then regular and one may simply expand the integrand in $\bar{s}$ around zero $e^{-\bar{s}}=1-\bar{s}+\dots$. This gives rise to terms linear in $\bar{s}$ in Eq.~(\ref{sp2}). In the region $K_d^\Lambda$ it is legitimate to use the asymptotic form of $\epsilon_{\bf k}$ [Eq.~(\ref{asymptotics})]. We then proceed along a line similar to Sec.~3.1. We perform the rescaling given by Eq.~(\ref{transf}) and pass to the spherical coordinates Eq.~(\ref{spherical}). It is then advantageous to specify $K_d^\Lambda$ in such a way that the transformation maps it onto a $d$-dimensional ball of a radius $\Lambda'$. One then performs the angular integrations and the asymptotic form of Eq.~(\ref{sp2})  yields
\begin{equation} 
-\frac{\bar{s}}{a\beta}+\frac{\mu}{a} = A_d\beta^{-1/\psi}\left[G_d\int_0^{\Lambda'} dr\frac{r^{2/\psi-1}}{e^{-\bar{s}}e^{r^2}-1}+c_0+c_1\bar{s}\dots \right]\;,
\label{sp3}
\end{equation}
where $A_d = \frac{1}{\pi^d}\left[\prod_{i=1}^{d}\left(\frac{1}{\alpha_i}t_i^{-1/\alpha_i}\right)\right]$, $G_d$ is a numerical constant arising from the angular integration, and, finally $c_0$ and $c_1$ are non-universal constants contributed by the  integration over the $\mathcal{R}^\Lambda$ region. The neglected terms are of order $\bar{s}^2$. We now observe that by extending the integration domain in Eq.~(\ref{sp3}) from $\Lambda'$ to infinity, we produce another contribution analytical in $\bar{s}$, i.e. 
\begin{equation} 
\int_0^{\Lambda'} dr\frac{r^{2/\psi-1}}{e^{-\bar{s}}e^{r^2}-1} = \int_0^{\infty} dr\frac{r^{2/\psi-1}}{e^{-\bar{s}}e^{r^2}-1}+d_0+d_1\bar{s}+\dots .
\end{equation}
The power series $d_0+d_1\bar{s}+\dots$ may be combined with $c_0+c_1\bar{s}\dots$ in Eq.~(\ref{sp3}). Subsequently we write
\begin{equation} 
\int_0^{\infty} dr\frac{r^{2/\psi-1}}{e^{-\bar{s}}e^{r^2}-1}=\frac{1}{2}\Gamma\left(\frac{1}{\psi}\right)g_{\frac{1}{\psi}}(e^{\bar{s}})\;,
\end{equation}
where we used the integral representation of the Bose functions
\begin{equation}
\label{sp4}
g_n(z)=\frac{1}{\Gamma(n)}\int_0^\infty dx\frac{x^{n-1}}{z^{-1}e^x -1 }\;.
\end{equation} 
For clarity we now rewrite Eq.~(\ref{sp3}) in the form 
\begin{equation}
 -\frac{\bar{s}}{a\beta}+\frac{\mu}{a}=B_d\beta^{-1/\psi}g_{1/\psi}(e^{\bar{s}})+C_d\beta^{-1/\psi}+E_d\beta^{-1/\psi}\bar{s}+\dots\;,
\label{sp5}
\end{equation}
where we incorporated all the non-universal ($T$-independent) constants into $B_d$, $C_d$ and $E_d$.
The asymptotic form of the Bose functions is given by \cite{Ziff77}
\begin{eqnarray}
\label{g_expansion}
g_{n}(e^{\bar{s}})\,-\,\zeta\left(n\right) \approx \left\{ 
\begin{array}{l l}
  \Gamma(1-n)\,|\bar{s}|^{n-1} & \quad \mbox{$1<n<2$} \\
  |\bar{s}|\,\log(|\bar{s}|) & \quad \mbox{$n=2$} \\ 
	-\,\zeta(n-1) \, |\bar{s}| & \quad \mbox{$n>2$}  \quad. \\ \end{array}  \right. 
\end{eqnarray}
Eq.~(\ref{sp5}) is structurally similar to the one studied for the continuum gas \cite{Napiorkowski13, Jakubczyk13}, the major difference being that the role of the dimensionality $d$ has now been taken over by the quantity $2/\psi$. The leading $\bar{s}$-dependent term in Eq.~(\ref{sp5}) is linear for $\frac{1}{\psi}>2$, and in consequence the dependence of $\bar{s}$ on $d$ drops out. The condition for the upper critical dimension $d_U$ therefore reads
\begin{equation}
\frac{1}{\psi}=2 \;,
\end{equation}
 which generalises the result $d_U=4$ \cite{Napiorkowski13} to systems characterised by an arbitrary dispersion admitted by Eq.~(\ref{asymptotics}). On the other hand, for $\frac{1}{\psi}<2$ the linear terms in Eq.~(\ref{sp5}) can be dropped.  
By plugging  the asymptotic formula of Eq.~(\ref{g_expansion}) into Eq.~(\ref{sp5}) we find the asymptotic solution for $\bar{s}$ in the vicinity of the phase transition, i.e. for $\epsilon=(\mu-\mu_c)/\mu_c\ll 1$ in the form 
 \begin{eqnarray}
\label{s_solution}
\bar{s} \sim \left\{ 
\begin{array}{l l}
  -\epsilon^{\psi/(1-\psi)} & \quad \mbox{$1<1/\psi<2$} \\
  \epsilon/\log(\epsilon) & \quad \mbox{$1/\psi=2$} \\ 
	- \epsilon & \quad \mbox{$1/\psi>2$}  \quad. \\ \end{array}  \right. 
\end{eqnarray}
By computing the grand canonical free energy $\omega$ from Eq.~(\ref{free_en}) it is possible to extract the exponent $\alpha$ defined by $\omega^{sing}\sim |\epsilon|^{2-\alpha}$. One obtains 
\begin{eqnarray}
\label{alpha}
\alpha = \left\{ 
\begin{array}{l l}
  \frac{1-2\psi}{1-\psi} & \quad \mbox{$1<1/\psi<2$} \\
	0 & \quad \mbox{$1/\psi>2$}  \quad. \\ \end{array}  \right. 
\end{eqnarray}
For the case $\alpha_i=2$ for $i\in\{1\dots d\}$ one recovers the exponents specific to the Berlin-Kac universality class \cite{Berlin52}. 

\section{Comparison to the ideal Bose gas}
In this Section, we analyse the ideal 
Bose gas with the dispersion asymptotics given by Eq.~(\ref{asymptotics}). This cannot
be done by taking $a\to 0$ since this operation does not commute with the thermodynamic limit.
The calculation is standard, so we quote only the key points.

The general expression for the grand partition function $\Xi$  of the Bose gas with the
one-particle dispersion $\epsilon_\kgr$ is given by \cite{Huang, Ziff77}
\[
\ln\Xi=-\sum_{\kgr\ne \0gr} \ln (1- z e^{-\beta \epsilon_\kgr} )- \ln(1-z),
\]
where $z=e^{\beta\mu}$ is determined by
\[
N=\sum_{\kgr\ne \0gr}\frac{z e^{-\beta \epsilon_\kgr} }{1-z e^{-\beta \epsilon_\kgr}}+ \frac{z}{1-z}\;.
\]
In the thermodynamic limit 
we have
\begin{eqnarray}
\left\{
\begin{array}{l}
\lim_\infty\frac{1}{V}\ln\Xi = -B_d(\beta, z),\\
n = b_d(\beta, z)+\lim\frac{1}{V}\frac{z}{1-z}.
\end{array}
\right.
\label{TD_IBG}
\end{eqnarray}
where:
\begin{equation}
B_d(\beta, z)= \frac{1}{(2\pi)^d}\int_{BZ}d^d k \ln (1- z e^{-\beta \epsilon_\kgr} )\;,
\label{Eq1IBG}
\end{equation}
\begin{equation}
b_d(\beta, z) = \frac{1}{(2\pi)^d} \int_{BZ}d^d k \frac{z e^{-\beta \epsilon_\kgr} }{1-z e^{-\beta \epsilon_\kgr}}\;.
\label{Eq2IBG}
\end{equation}
The phase diagram of the ideal Bose gas has little in common with that of the IBG, since is not defined for $\mu>0$, and condensation may occur only for $\mu=0$.
The present analysis proceeds analogously to the case
of the standard continuum model  \cite{Huang, Ziff77} and we here list only the aspects of present relevance.

$\bullet$ The Bose-Einstein condensation is present if the integral (\ref{Eq2IBG}) is convergent for $z=1$, otherwise there is no phase transition.
Observe that the convergence condition for $b_d(\beta, 1)$ is identical as for the integral of Eq.~($\ref{I_d})$ for $\bar{s}=0$. 
 In consequence the system displays Bose-Einstein condensation exclusively for $\psi < 1$  
which is the same as for the IBG. Note however, that the quantity $\psi$ cannot be given the interpretation in terms of the shift exponent (which now has no sense at all), since condensation is restricted to $\mu=0$. 

$\bullet$ The critical exponents of these two models do not coincide. Let us exemplify this with the behaviour of 
the specific heat $C$. For the noninteracting Bose gas it turns out to behave as in the continuum model in $d$ dimensions \cite{Ziff77, May}, where however
the dimension $d$ of the continuum model is replaced by $d_{ef}=\frac{2}{\psi}$. We obtain the following behaviour
near the critical temperature ($n$ is fixed and we 
consider the temperature dependence):
\begin{enumerate}
\item The specific heat is a continuous (although non-smooth) function of temperature 
for $d_{ef}<4$. The left- and right-side derivatives at the
critical point are finite.
 \item For $d_{ef}=4$, the specific heat is also continuous at $T_c$, but the right-hand derivative is infinite.
\item For $d_{ef}>4$ there is a discontinuity in the specific heat itself. Note that such a feature is characteristic to Landau theory.
\end{enumerate}
The asymptotic behaviour of $b(\beta, z)$ for $z\approx 1$ does not enter into the behaviour or the specific heat --
only the convergence properties of $b(\beta, z)$ are important.

$\bullet$ The asymptotic behaviour of $b(\beta, z)$ for $z\approx 1$ however influences the behaviour of specific heat near $T=0$.
It turns out that we have the following asymptotics:
\begin{equation}
C\sim T^{\frac{1}{\psi}}.
\end{equation}
In $d=3$, for the standard quadratic asymptotics of $\epsilon_\kgr$ we recover the behaviour characteristic to the continuum model:
$C\sim T^{3/2}$, and for $\alpha_1=\alpha_2=\alpha_3=1$, we have: $C\sim T^3$, i.e. the same behaviour as for photons in a continuum 
model.

\section{Correlation function}
In this section we analyse the correlation function of the anisotropic imperfect Bose gas defined in Sec.~2. As was shown in Ref.~\cite{Napiorkowski12} for the isotropic case, this can be obtained from the correlation function of the noninteracting Bose gas upon replacing the chemical potential $\mu$ with $\tilde{\mu} = \mu - a n(T,\mu)$. The derivation of this result is not sensitive to the particular form of the dispersion and straightforwardly carries over to the present, more general case. The parameter $\tilde{\mu}(T,\mu)$ is obtained from 
\begin{equation}
n(T,\mu)= -\frac{\partial \omega}{\partial\mu}  =\frac{\mu}{a}-\beta^{-1}\frac{\partial}{\partial \mu}\phi\left(\bar{s}(T,\mu),T,\mu\right),
\end{equation}
which leads to 
\begin{equation}
n(T,\mu)=\frac{\mu}{a}-\frac{\bar{s}}{a\beta}\;,
\end{equation}
and, in consequence 
\begin{equation}
\tilde{\mu}=\beta^{-1}\bar{s}\;.
\end{equation}
This relation may be viewed as an interpretation of the parameter $\bar{s}$ in terms of a renormalised chemical potential.
With the above knowledge, the problem now becomes reduced to an analysis of the correlation function of the anisotropic perfect Bose gas, followed by the replacement $\mu\longrightarrow \tilde{\mu}=\beta^{-1}\bar{s}$ and an extraction of the physically meaningful quantities with particular focus on the transition at $\bar{s}\to 0^-$. 

Clearly, the exponent governing the singularity of the correlation length   at the phase transition depends on the direction. What is less obvious, also the presence (or absence) of oscillations of the correlation function $\chi({\bf x})$ as function of the distance $|{\bf x}|$ (at $T>T_c$) does depend on the direction of ${\bf x}$ relative to the anisotropies of the dispersion.  
\subsection{Correlation function of the perfect anisotropic Bose gas: asymptotic behaviour} 
The density-density correlation function is defined by 
\begin{equation}
\chi({\bf x}_1,{\bf x}_2)=n_2({\bf x}_1,{\bf x}_2)-n^2\;,
\end{equation}
where the two-particle density $n_2({\bf x}_1,{\bf x}_2)$ is related to the 2-particle (grand canonical) density matrix $\rho_2({\bf x}_1, {\bf x}_2, {\bf x}_1', {\bf x}_2')$ by
\begin{equation}
n_2({\bf x}_1,{\bf x}_2) = \rho_2({\bf x}_1, {\bf x}_2, {\bf x}_1, {\bf x}_2)\;,
\end{equation}
and $n$ denotes the density.
 For noninteracting bosons one finds (see e.g.\cite{Ziff77})
\begin{equation}
\rho_2({\bf x}_1, {\bf x}_2, {\bf x}_1', {\bf x}_2')= \rho_1({\bf x}_1, {\bf x}_1')\rho_1({\bf x}_2, {\bf x}_2')+\rho_1({\bf x}_2, {\bf x}_1') \rho_1({\bf x}_1, {\bf x}_2')\;,
\end{equation} 
where the one-particle density matrix $\rho_1({\bf x}_1, {\bf x}_2)$ is in the thermodynamic limit given by 
\begin{equation}
\rho_1({\bf x}_1, {\bf x}_1')=\int\frac{d^d k}{(2\pi)^d}\rho_{\bf k}e^{i{\bf k}({\bf x}_1-{\bf x}_1')}\;, 
\label{rhodef}
\end{equation}
and therefore depends on ${\bf x} ={\bf x}_1- {\bf x}_1'$. Here 
\begin{equation}
\rho_{\bf k} = \left(z^{-1}e^{\beta\epsilon_{\bf k}}-1\right)^{-1}= \sum_{j=1}^{\infty}z^je^{-\beta j\epsilon_{\bf k}}
\label{BEdistr}
\end{equation}
is the Bose-Einstein distribution.
Using the above, we evaluate $n_2({\bf x}_1,{\bf x}_2)=\rho_2({\bf x}_1, {\bf x}_2, {\bf x}_1, {\bf x}_2)$ and find 
\begin{equation}
n_2({\bf x}_1,{\bf x}_2)= \rho_1^2({\bf x}_1,{\bf x}_2)+n^2
\end{equation}
In consequence:
\begin{equation}
\chi({\bf x}_1,{\bf x}_2)=\chi({\bf x}_1-{\bf x}_2)=\rho_1^2({\bf x}_1-{\bf x}_2)=\rho_1^2({\bf x})\;.
\end{equation}
In what follows, we denote $\rho_1({\bf x})$ as $\rho({\bf x})$, so that $\chi({\bf x})= \rho({\bf x})^2$. 

For $|{\bf x}|\to \infty$ the summation over $j$ may be replaced by an integration over a continuous variable $\sum_j\longrightarrow \int d\tilde{t}$. The subsequent change of variables $t=\frac{\tilde{t}}{x}$ brings $\rho({\bf x})$ [given by Eq.~(\ref{rhodef})] to the following form 
\begin{equation}
\rho({\bf x}) \simeq \int_{\mathbf{R}^d}\frac{d^dk}{(2\pi)^d}\int_0^\infty dt xe^{x\Phi({\bf k}, t)}\;,
\end{equation}
where 
\begin{equation}
\Phi({\bf k}, t) = - \beta t \epsilon_{\bf k} +i\frac{{\bf k}{\bf x}}{x}+\beta\mu t\;.
\end{equation}
The asymptotic behaviour of $\rho({\bf x})$ for $x\to\infty$ can now be extracted with the steepest-descent method. Evaluating the derivatives of $\Phi({\bf k}, t)$ yields the saddle-point condition 
\begin{eqnarray}
\left\{ 
\begin{array}{l l}
\label{sp_corr}
  \epsilon_{{\bf k}_0}=\mu    \\
  \beta t_0\partial_{k_j}\epsilon_{\bf k}|_{{\bf k}_0} -	i\frac{x_j}{x} = 0  & \quad \mbox{$ j \in \{1\dots d\} $}   \\ \end{array}  \right. 
\end{eqnarray}
for $t_0$, ${\bf k}_0$.
Below we solve Eq.(\ref{sp_corr}) for specific cases and evaluate the asymptotic form of the correlation function. 
\subsubsection{Case $\epsilon_{\bf k}= \frac{1}{2} \sum_{j=1}^{d-1}k_j^2+\frac{\tau}{4}k_d^4$ and ${\bf x}=x{\bf e}_1$}
\quad \\
In this case ${\bf x}$ is taken perpendicular to the anisotropy. Eq.~(\ref{sp_corr}) then yields the solution 
\begin{equation}
{\bf k}_0 = i\sqrt{-2\mu}{\bf e}_1\;, \;\;\;\; t_0 = \beta^{-1}(-2\mu)^{-\frac{1}{2}}
\end{equation}
with $\Re(t_0)>0$.
From the Laplace formula one now obtains 
\begin{equation} 
\label{rho1}
\rho({\bf x}) \sim e^{x\Phi({\bf k}_0, t_0)} = e^{-x/2\xi\perp}\;,
\end{equation}
which allows us to identify the correlation length $\xi_\perp=[\sqrt{-2\mu}/2]^{-1}$ together with the critical exponent $\nu_\perp= \frac{1}{2}$, which is precisely the same as for the standard isotropic case.
\subsubsection{Case $\epsilon_{\bf k}= \frac{1}{2} \sum_{j=1}^{d-1}k_j^2+\frac{\tau}{4}k_d^4$ and ${\bf x}=x{\bf e}_d$}
\quad \\
Taking ${\bf x}$ along the anisotropy, we obtain the following two solutions to Eq.~(\ref{sp_corr}) with $\Re (t_0)>0$:
\begin{equation}
{\bf k}_0^{(1)} = \left(\frac{-4\mu}{\tau}\right)^{1/4}e^{\frac{5}{4}\pi i} {\bf e}_d\;, \;\;\;\; t_0^{(1)}= \frac{1}{\beta \tau}\left(\frac{-4\mu}{\tau}\right)^{-3/4}e^{\frac{\pi i}{4}}\;,
\end{equation}
\begin{equation}
{\bf k}_0^{(2)} = \left(\frac{-4\mu}{\tau}\right)^{1/4}e^{\frac{7}{4}\pi i} {\bf e}_d\;, \;\;\;\;  t_0^{(2)}= \frac{1}{\beta \tau}\left(\frac{-4\mu}{\tau}\right)^{-3/4}e^{-\frac{\pi i}{4}}\;.
\end{equation}
In consequence, 
\begin{equation}
x\Phi({\bf k}_0^{(l)}, t_0^{(l)})=\frac{1}{\sqrt{2}}\left(-1\pm i\right)(\frac{-4\mu}{\tau})^{1/4}x\quad \textrm{ for} \quad l\in \{1,2\}\;.
\end{equation}
The moduli of both contributions are equal and therefore both points contribute to the density matrix. We find 
\begin{equation}
\rho({\bf x})\sim e^{x\Phi({\bf k}_0^{(1)}, t_0)} + e^{x\Phi({\bf k}_0^{(2)}, t_0)} = 2 \cos \left[\left(\frac{-\mu}{\tau}\right)^{1/4}x\right]e^{-\left(\frac{-\mu}{\tau}\right)^{1/4}x}\;.
\end{equation}
This allows us to read off the exponent $\nu_\parallel = \frac{1}{4}$. We observe the oscillatory behaviour of $\rho({\bf x})$, which may be contrasted with the case ${\bf x}=x {\bf e}_1$ [Eq.~(\ref{rho1})], where we found monotonous decay. 
\subsubsection{Case $\epsilon_{\bf k}= \frac{1}{2} \sum_{j=1}^{d-1}k_j^2+\frac{\tau}{4}k_d^{2m}$  ($m\in \mathbf{N}$)}
\label{subsubsec:WyklNu}
\quad \\
The analysis of the previous section may be generalised to a situation where $\epsilon_{\bf k}= \frac{1}{2} \sum_{j=1}^{d-1}k_j^2+\frac{\tau}{4}k_d^{2m}$ and $m>1$ is a natural number. For the 'radial' case (${\bf x}= x{\bf e}_1$) we obtain a monotonous behaviour of $\rho({\bf x})$ with the exponent $\nu_\perp =\frac{1}{2}$. For the 'axial' case (${\bf x}= x{\bf e}_d$) one finds oscillatory behaviour of $\rho({\bf x})$ and $\nu_\parallel = \frac{1}{2m}$.
\subsection{Correlation function of the perfect anisotropic Bose gas: general expressions} 
Above we extracted the asymptotic behaviour of $\rho({\bf x})$ at large $|{\bf x}|$ in the 'radial' and 'axial' directions. For $\epsilon_{\bf k}= \frac{1}{2} \sum_{j=1}^{d-1}k_j^2+\frac{\tau}{4}k_d^{4}$ one may however also obtain an exact expression for $\rho({\bf x})$ valid for any ${\bf x}$. This is achieved by factorising the integral in Eq.~(\ref{rhodef}) and directly evaluating the expression. One finds
\begin{equation}
\rho({\bf x})
=
 \sum_{j=1}^\infty z^j 
\prod_{\alpha=1}^{d-1} F_2(x_\alpha, \beta, j) F_4(x_d, \beta, j)\;,
\label{rho2...24}
\end{equation}
where 
\begin{equation}
F_2(x_\alpha, \beta, j) = 
\sqrt{\frac{2\pi}{\beta j}}
\exp\left(-\frac{x_\alpha^2}{2 \beta j}\right);
\label{TrF2}
\end{equation}
and
\begin{equation}
F_4(x, \beta, j)
=
\frac{\pi  \left(\frac{\sqrt{{\beta} j} \, _0F_2\left(;\frac{1}{2},\frac{3}{4};\frac{x^4}{64 {\beta} j}\right)}{\Gamma \left(\frac{3}{4}\right)}-\frac{x^2 \, _0F_2\left(;\frac{5}{4},\frac{3}{2};\frac{x^4}{64 {\beta} j}\right)}{\Gamma \left(\frac{1}{4}\right)}\right)}{({\beta} j)^{3\slash{}4}}\;.
\end{equation}
Above we put $\tau=1$.
The symbol $_0F_2\left(;b_1,b_2;z\right)$ denotes  one of the {\em generalised hypergeometric functions}. The generalised hypergeometric function is a function depending of  $p+q$ parameters $a_1,\dots,a_p$, $b_1,\dots,b_q$ and the variable $z$. Its  definition reads \cite{Slater}, \cite{Wolfram}
\[
_pF_q\left(a_1,\dots, a_p;b_1,\dots, b_q;z\right)
=
\sum_{k=0}^\infty \frac{(a_1)_k \dots (a_p)_k}{(b_1)_k\dots(b_q)_k}
\frac{1}{k!} z^k \;.
\]
In the above formula, $(a)_k$ denotes the Pochhammer symbol.
 
   Despite being exact, Eq.~(\ref{rho2...24}) is inconvenient for extracting the asymptotic expressions derived in the previous section. It may however be easily plotted for arbitrary choices of ${\bf x}$ without recourse to asymptotic expansions. This is done for $d=3$ in Figures 1-3. 
\begin{figure}
\begin{center}
 \includegraphics[width=12cm]{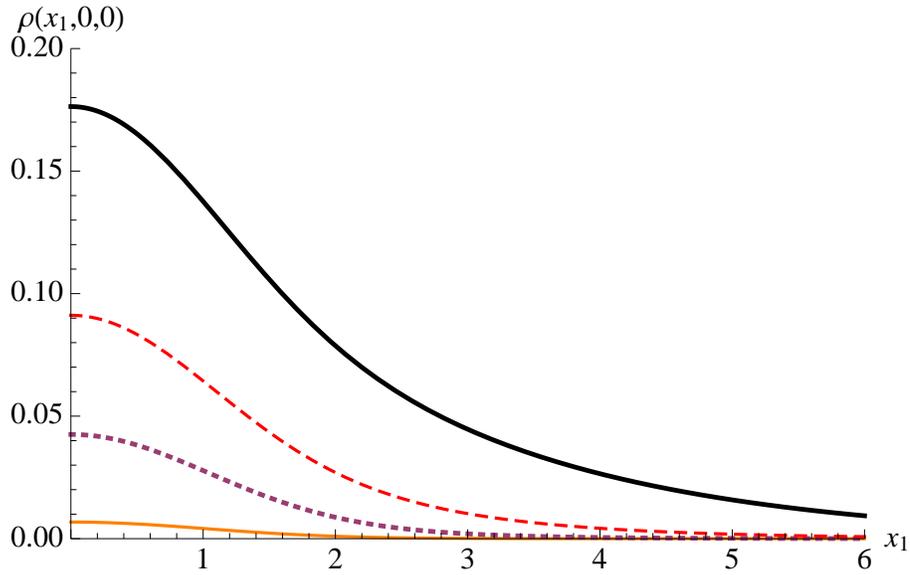}
\end{center}
\caption{(color online) The density matrix $\rho({\bf x})$ in the radial direction (for ${\bf x}=x {\bf e}_1$) for a sequence of $z$ approaching Bose-Einstein condensation. The plotted curves correspond to $\beta=1$ and $z=0.1$ (the lowest curve), $z=0.5,\;z=0.8$, $z=1$ (the highest curve). The quantity $\rho({\bf x})$ decreases monotonously in agreement with the results of Sec. 5.1.1.    }
\end{figure}  
\begin{figure}
\begin{center}
 \includegraphics[width=12cm]{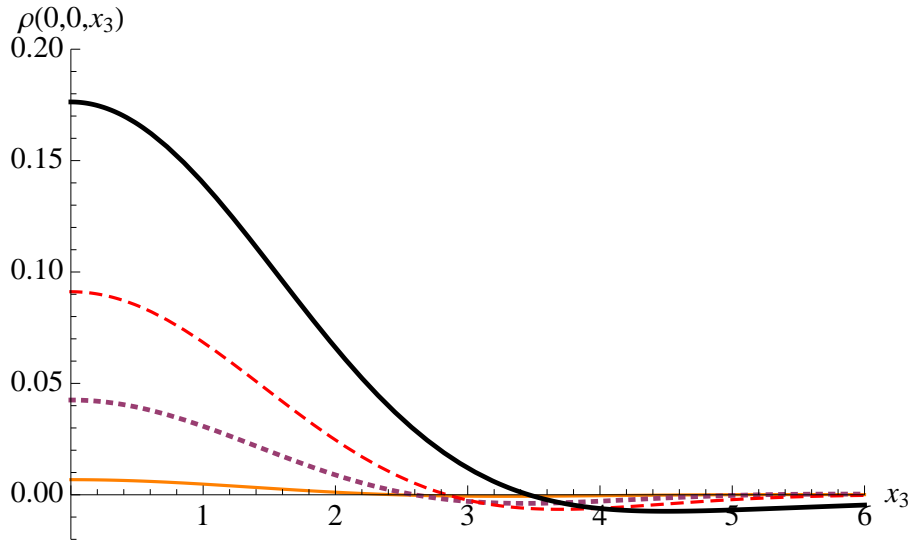} 
\end{center}
\caption{(color online) The density matrix $\rho({\bf x})$ in the axial direction (for ${\bf x}=x {\bf e}_3$) for a sequence of $z$ approaching Bose-Einstein condensation. The plotted curves correspond to $\beta=1$ and $z=0.1$ (the lowest curve), $z=0.5,\;z=0.8$, $z=1$ (the highest curve). The quantity $\rho({\bf x})$ shows oscillatory behaviour in agreement with the results of Sec. 5.1.2.  }
\end{figure}  
\begin{figure}
\begin{center}
 \includegraphics[width=12cm]{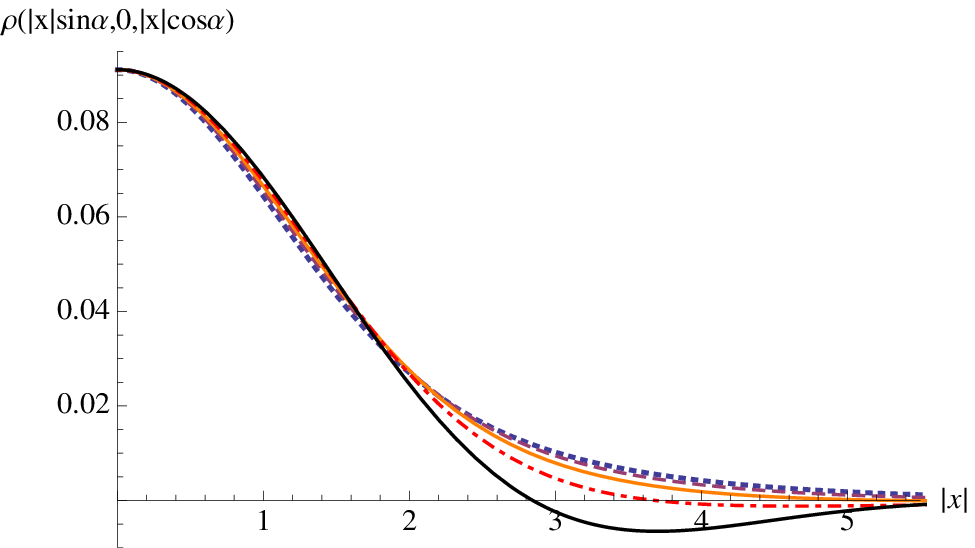} 
\end{center}
\caption{(color online) Evolution of $\rho({\bf x})$ upon varying the direction of ${\bf x}=|x|\sin\alpha {\bf e}_1 + |x|\cos\alpha {\bf e}_3$. The plotted curves correspond to $\beta=1$, $z=0.8$, and $\alpha=0$ (the bold curve reaching the lowest-lying minimum), $\alpha=\pi/6$, $\alpha=\pi/4$, $\alpha=\pi/3$, $\alpha=\pi/2$ (the dotted, monotonous curve).  }
\end{figure}  
The resulting plots confirm the monotonous character of $\rho({\bf x})$ in the radial direction (${\bf x}=x {\bf e}_1$) and the occurrence of damped oscillations in the axial direction (${\bf x}=x {\bf e}_3$). More generally, oscillations occur for ${\bf x}$ such that ${\bf x}{\bf e}_3\neq 0$, as can be seen  in Figure 3. 

\subsection{Correlation function of the imperfect anisotropic Bose gas} 
As explained at the beginning of this  section, the correlation function of the imperfect Bose gas can now be obtained by performing the replacement $\mu\longrightarrow \tilde{\mu}=\beta^{-1}\bar{s}$ in the results obtained above in the absence of any interactions. We find   
 \begin{eqnarray}
\label{nu_i}
\nu_{\perp /\parallel}^{(IBG)} = \left\{ 
\begin{array}{l l}
  \frac{\psi}{1-\psi}\nu_{\perp /\parallel} & \quad \mbox{$1<1/\psi<2$} \\
	\nu_{\perp /\parallel} & \quad \mbox{$1/\psi>2$}   \\ \end{array}  \right. 
\end{eqnarray} 
as the exponents controlling the divergence of the correlation lengths at Bose-Einstein condensation. The exponents $\nu_\perp$ and $\nu_\parallel$ were obtained explicitly in the  Sec.~ \ref{subsubsec:WyklNu} 
for the case of  $\epsilon_{\bf k}= \frac{1}{2} \sum_{j=1}^{d-1}k_j^2+\frac{\tau}{4}k_d^{2m}$, with integer $m$. Note that Eq.~(\ref{nu_i}) holds in the high-temperature phase.

For the special case of $\alpha_i=2$ for $i\in\{1\dots d\}$ one recovers $\nu=\frac{1}{d-2}$ below the upper critical dimension, which then equals 4. 

\section{Remarks on the relation to the spherical model} 
The classical spherical model \cite{Berlin52, Joyce} is among the  most recognised statistical physics models that display a phase transition, and, at the same time, admit an analytical solution. As already remarked, the IBG and the spherical model fall into the same bulk universality class at $T>0$, which is also specific to the $\mathcal{N}\to \infty$ limit of $O(\mathcal{N})$-symmetric models \cite{Stanley68, Pearce77}. In addition to the most studied cases (such as only nearest-neighbour interactions) leading to the standard Berlin-Kac exponents, one may manipulate the model couplings $J_{ij}$ (competing ferro- and antiferromagnetic interactions) and obtain a picture similar to the one described in Sec.~3 for the IBG \cite{Selke77, W96}.     

Despite this above affinity between the two models we would like to point out a difference. While the IBG Hamiltonian contains a kinetic energy contribution, the spherical model does not. As a result, the IBG is naturally equipped with two parameters ($T$ and $\mu$) to tune the transition and is well defined for $T\to 0 $. In consequence, its phase diagram realises the standard scenario of quantum criticality for $T$ small \cite{Jakubczyk13, Jakubczyk15}. This is not quite the case for the Berlin-Kac model, where a realisation of quantum criticality requires defining its quantum extension, which in practice usually amounts to adding an extra kinetic term to the Hamiltonian. This procedure is however not unique and leads to several different universality classes for $T$ approaching zero, which, for example, may be characterised by distinct $\psi$ exponents. Different variants of the quantum spherical model have been addressed in Refs.~\cite{VZ92, Nieuwenhuizen95, Vojta96, Chamati98, Gracia04, Menezes08, Bienzobar13, Wald15}. For some of them, the non-standard critical behaviour was observed for fine-tuned coupling constants \cite{VCW97, Gomes12}. 

Another interesting connection between the IBG and the spherical model occurs via the Casimir forces. Ref.~\cite{Napiorkowski13} found the Casimir scaling functions different by a factor of 2 from those of the spherical model. An explanation for this fact was recently given in Ref.~\cite{Diehl17}. The anisotropies of the correlation functions should find a reflection in the form of the scaling functions for the Casimir energy, which may be an interesting direction for future studies.  


\section{Summary}
Realising the imperfect Bose gas on a lattice allows for engineering the kinetic couplings in such a way that the asymptotic behaviour of the dispersion is altered \cite{KwartyczneChinczyki}.  This influences the universality class of Bose-Einstein condensation. By means of an exact analysis of the considered model, we have argued that the universality class is only indirectly related to the system dimensionality. Instead, it is fully specified by the shift exponent $\psi$. This in turn is given by Eq.~(\ref{psi}), and, for the present system, is not determined exclusively by the system dimensionality $d$ and the dynamical exponent $z$. The derived condition for the occurrence of the Bose-Einstein condensate (at arbitrarily low but finite $T$) reads $1/\psi>1$. Non-classical critical behaviour occurs for $1/\psi\in ]1,2[$. The universality class is then the same as that of the classical spherical model, albeit in dimensionality $d_{ef}=\frac{2}{\psi}$. For $1/\psi>2$ Bose-Einstein condensation is characterised by classical (Landau) critical exponents. Our analysis indicates (both the perfect and imperfect Bose gases) a qualitative change in the behaviour of the correlation function introduced by the anisotropies.  The character of the divergence of the correlation length acquires a dependence on direction (the exponents $\nu_\perp$ and $\nu_\parallel$ are of different values). The relation between these exponents in the cases of perfect and imperfect Bose gases is linear with a coefficient specified by the shift exponent $\psi$ [Eq.~(\ref{nu_i})]. The correlation function may display either a monotonous decay, or exponentially damped oscillations. The latter is found more generic in the simplest case of $\sim k^2$ dispersion in all directions but one. This is in contrast to the standard isotropic case with $\epsilon_{\bf k}\sim k^2 $ where the oscillations are absent. 

Our result for the correlation function poses a natural question whether similar effects (presence or absence of oscillations of $\rho({\bf x})$ in anisotropic Bose gases) occur also for interacting systems. It would also be interesting to investigate the finite-size effects in systems with such anisotropies, in particular see if and how the direction dependence of the correlation function and its oscillatory behaviour transfers to the scaling functions for the Casimir energy.

\ack
We are grateful to Walter Metzner and Jaromir Panas for reading the manuscript and a number of useful comments. We thank Marek Napi\'{o}rkowski for  useful discussions.  
PJ acknowledges support from the Polish National Science Center via grant 2014/15/B/ST3/02212.

\end{document}